\documentclass[12pt,preprint]{aastex}

\def\ltsima{$\; \buildrel < \over \sim \;$}
\def\gtsima{$\; \buildrel > \over \sim \;$}
\def\lsim{\lower.5ex\hbox{\ltsima}}
\def\gsim{\lower.5ex\hbox{\gtsima}}
\def\psr{PSR~J1740-5340}
\def\com{COM~J1740-5340}
\def\Msun{${\rm M}_{\odot}$}

\begin{document}

\title{New clues on the nature of the companion to PSR~J1740-5340 in
  NGC6397 from XSHOOTER spectroscopy. \footnote{Based on data taken at
  the ESO, within the observing programs 085.D-0377(A) and
  087.D-0716(A).}}

\author{A. Mucciarelli$^{1}$, M. Salaris$^{2}$, B. Lanzoni$^{1}$, C. Pallanca$^{1}$, 
E. Dalessandro$^{1}$, F. R. Ferraro$^{1}$}

\affil{$^{1}$Dipartimento di Fisica \& Astronomia, Universit\`a 
degli Studi di Bologna, Viale Berti Pichat, 6/2 - 40127
Bologna, ITALY}

\affil{$^{2}$Astrophysics Research Institute, Liverpool John Moores University, 
    12 Quays House, Birkenhead, CH41 1LD, United Kingdom }

\begin{abstract}  
By using XSHOOTER spectra acquired at the ESO Very Large Telescope, we
have studied the surface chemical composition of the companion star to
the binary millisecond pulsar \psr\ in the globular cluster NGC
6397. The measured abundances of Fe, Mg, Al and Na confirm that the
star belongs to the cluster.
On the other hand, the measured surface abundance of nitrogen
([N/Fe]$=+0.53\pm 0.15$ dex) combined with the carbon upper limit
([C/Fe]$<-2$ dex) previously obtained from UVES spectra allow us to
put severe constraints on its nature, strongly suggesting that the
pulsar companion is a deeply peeled star.  In fact, the comparison
with theoretical stellar models indicates that the matter currently
observed at the surface of this star has been processed by the
hydrogen-burning CN-cycle at equilibrium. In turn, this evidence
suggests that the pulsar companion is a low mass ($\sim 0.2$ \Msun)
remnant star, descending from a $\sim 0.8$ \Msun\ progenitor which
lost $\sim 70-80$\% of its original material because of mass transfer
activity onto the pulsar.
\end{abstract}  
 

\section{Introduction}   
\label{intro}  
The dense and dynamically active environment typical of Galactic
globular clusters (GCs) provide the ideal conditions for the formation
and evolution of stellar exotica, such as blue straggler stars,
interacting binaries and millisecond pulsars (hereafter MSPs; see
\citealp{bai95,bella95,ferraro95,ferraro09,ferraro12}). In particular
MSPs are generated in binary systems containing a neutron star, which
is eventually spun-up through mass accretion from an evolving
companion. The final state therefore is a deeply peeled or even
exhausted star (as a white dwarf) orbiting a rapidly spinning pulsar.

The MSP J1749-5340, discovered in the GC NGC 6397
\citep{damico01a}, belongs to a binary system with a  orbital period of $\sim
1.35$ days. At 1.4 GHz  radio frequency it shows eclipses for about 40\% of the orbital
period, likely due to matter released from the companion
\citep{damico01b}, which probably is also responsible for its X-ray
emission \citep[see][]{grindlay01}.
The companion star (hereafter, \com) was identified by \citet{ferraro01}
as a variable star with a luminosity comparable to that of the Main
Sequence (MS) Turnoff and an anomalously redder color (see right
panel in Figure \ref{loca}).  The shape of its light curve suggests
that it is tidally distorted by the interaction with the pulsar
\citep{ferraro01, kaluzny03}.

In virtue of its brightness ($V\sim16.6$), \com\ represents one of the
rare cases where spectroscopy of a GC MSP companion can be successfully
performed, accurately investigating its kinematical and chemical properties. 
Indeed, a detailed study of the companion
radial velocity curve was performed by \citet{ferraro03} and, once
combined with the PSR radial velocity curve, it allowed to derive the
mass ratio of the system ($M_{\rm PSR}/M_{\rm COM}=5.85\pm 0.13$). In
turn, this constrained the companion mass in the range $0.22$
\Msun$\le M_{\rm COM}\le 0.32$ \Msun \citep{ferraro03,kaluzny03}. In
addition, also a chemical analysis has been performed, highlighting
{\sl (a)} a complex structure of the H$\alpha$ profile, well
reproduced by two different emission components \citep{sabbi03a}, {\sl
  (b)} the unexpected detection of a He I line, suggesting the
existence of a hot (T$>$10000 K) region located on the stellar
hemisphere facing the MSP \citep{ferraro03}, and {\sl (c)} some
anomalous chemical patterns (for Li, Ca and C) with respect to the
GC chemical composition \citep{sabbi03b}.

Several hypothesis have been proposed to explain the nature of this
system and find a coherent picture for the observational evidence
collected \citep{possenti,burderi02,orosz}.  In
particular, two possible origins for \com\ can be advanced: {\sl (1)}
it is a low-mass ($<0.3$ \Msun), MS star perturbed by the pulsar; {\sl
  (2)} it is a {\sl normal} star (at the Turnoff or slightly evolved,
according to its luminosity), deeply peeled by mass loss down to the
present mass.

Following the suggestions of \citet{ergma03}, the CN surface
abundances are an ideal tool to discriminate between the two proposed
scenarios: in fact, if it is a perturbed
MS star, its C and N abundances will be unmodified with respect to the
pristine cluster chemical composition. On the other hand, if it is a
peeled star, its chemical composition will show the signatures of
H-burning CN-cycle (in particular, a decrease of $^{12}$C 
and an increase of $^{14}$N)
\footnote{Note that this is also the
  chemical signature used by \citet{ferraro06} to infer a
  mass-transfer origin for a sub-sample of blue straggler stars in 47
 Tucanae.}.
A first evidence in favour of the latter scenario has been
provided by the significant lack of C in \com,
through the analysis of UVES@VLT high-resolution spectra \citep{sabbi03b}. 
Unfortunately, however, those spectra do not
allow to measure the N abundance. 
In this paper we present XSHOOTER spectroscopic observations of \com, 
focussing on the N abundance. 

\section{Observations}
Observations of \com\ were secured with the XSHOOTER spectrograph at
the ESO-VLT. A second target (hereafter MS1) was included in the same
slit and used as comparison star. This is a MS star with
$V_{606}=17.28$, located at $\sim 1.3\arcsec$ from the main target
(see Figure \ref{loca}, right panel; two other objects lie within the slit,
their faintness, $V_{606}>19.5$).  
A first observing run has been performed in June 2010, enabling
simultaneously the UVB ($\sim$ 3300-5500 \AA) and the VIS
($\sim$5500-10000 \AA) channels of XSHOOTER.  The adopted slit width
was $0.8\arcsec$ (R=6200) and $0.7\arcsec$ (R=11000) for the UBV and
VIS channels, respectively, and the exposure time was 1200 s in both
cases. To increase the SNR in the region around the NH band
($\sim$3360 \AA), a second observation has been secured in July 2011,
using only the UVB channel, with the same slit width and with an
exposure time of 2700 s.

The data reduction was performed with the XSHOOTER ESO pipeline, version
2.0.0, including bias subtraction, flat-fielding, wavelength calibration, 
rectification and order merging.  Because the pipeline does not support efficiently
the spectral extraction for many sources in the same slit, this task
was performed manually with the IRAF package {\sl apall} in optimal
extraction mode.  The final spectra have SNR=50-100 for COM
J1740-5340, and SNR=40-70 for MS1.

\section{Chemical analysis}

The chemical abundances of Fe, C, N, Na, Al and Mg have been derived
through a $\chi^2$-minimization between the observed spectral features
and a grid of synthetic spectra computed with different abundances for
each species, following the procedure described in \citet{m12b}.  With
respect to the traditional method of the equivalent widths, this
approach reduces the difficulties in the continuum location, a
critical task in the analysis of low-resolution spectra because of the
severe line blanketing conditions.

For the analysis of \com\ we adopted the atmospheric parameters
($T_{eff}$=~5530 K, log~g=~3.46 and $v_{turb}$=~1.0 km/s) derived by \citet{sabbi03b}.  
For MS1 we derived  $T_{eff}$= 6459 K and $\log g=4.44$ by comparing the position 
of the star in the CMD with a theoretical isochrone from the BaSTI dataset 
\citep{pietr06} with Z=0.0003 
(corresponding to [Fe/H]=--2.1),
$\alpha$-enhanced chemical mixture, and an age of 12 Gyr, assuming 
the reddening and the distance modulus by \citet{f99}. 
The photometric $T_{eff}$ is confirmed by
the analysis of the wings of the H$\alpha$ line 
(the $T_{eff}$ of \com\ has been derived from the 
H$\alpha$ wings, thus we can considered the $T_{eff}$ of the two objects 
on the same scale). 
For the microturbulent velocity 
we assumed 1 km s$^{-1}$, that is a
reasonable value for unevolved low-mass stars \citep{gratton01}.

The synthetic spectra have been computed with the SYNTHE code by
R. L. Kurucz \citep{sbordone}, including all the
atomic and molecular transitions listed in the Kurucz/Castelli 
line list\footnote{http://wwwuser.oat.ts.astro.it/castelli/linelists.html}.
All the synthetic spectra have been convolved with a Gaussian profile
to reproduce the appropriate spectral resolution.
The synthetic spectra used for the analysis of COM J1740-5340 have been
also convolved with a rotational profile with $v \sin i$=~50 km s$^{-1}$, 
\citep{sabbi03b}.  Instead, for MS1 no additional rotational
velocity is added, according to the very low values ($<$3-4 km
s$^{-1}$) typically measured in unevolved low-mass stars
\citep{lucatello03}.  The model atmospheres have been calculated with
the ATLAS9 code \citep{castelli04}, 
assuming with [M/H]$=-2.0$ dex and $\alpha$-enhanced
chemical composition \citep[according to the analysis by][]{sabbi03b}.

The spectral lines for the analysis have been selected
through the detailed inspection of the synthetic spectra, considering
only those transitions that are unblended at the XSHOOTER resolution.
A total of 15
and 13 Fe~I lines have been selected in COM~J1740-5340 and in MS1, respectively.
The nitrogen abundances were derived by fitting the band-head of the
A-X (0-0) and (1-1) transitions located at 3360 \AA\ and 3370 \AA,
respectively. 
The inspection of the solar-flux spectrum 
by \citet{neckel} suggested that we need to decrease by 0.5 dex the
Kurucz $\log$ gf, in order to properly reproduce the solar N
abundance. 
The carbon abundances were derived from the CH G-band at 4300 \AA.
The Kurucz $\log$ gf for the CH transitions were decreased by 0.3 dex,
in order to reproduce the G-band observed in the solar-flux spectrum
by \citet{neckel},
as discussed in \citet{m12}. 
Aluminum abundances were derived from the UV resonance line at 3961
\AA, applying a non-LTE correction of +0.7 dex for both targets,
according to the calculations of \citet{andr08}.
To derive the sodium abundance, we used the Na doublet at 8183-8194
\AA: these lines fall in a spectral region severely contaminated by
telluric features. Despite the accuracy of the telluric subtraction
(performed with the IRAF task {\sl telluric} by adopting as template
the spectrum of an early-type star observed during the observing runs),
the radial velocities of
the two stars prevent a total deblending between the Na lines and the
telluric features.  For both stars we therefore provide only upper
limits for the Na abundance, including the non-LTE corrections by
\citet{lind}. 

Abundance uncertainties have been estimated by adding in quadrature
the errors obtained from the fitting procedure and those arising from
the atmospheric parameters.  The uncertainties in the fitting
procedure have been estimated by resorting to MonteCarlo simulations.
Uncertainties due to atmospheric parameters are calculated by varying 
one parameter at a time 
while keeping the other ones fixed, and repeating the analysis.

\section{Results}
Table 1 lists the chemical abundances derived for \com\ and MS1,
together with their total uncertainties.  The iron content of MSP
companion is [Fe/H]$=-2.00\pm 0.12$ dex, in agreement with both the
iron abundance of MS1 ([Fe/H]$=-1.93\pm 0.18$ dex) and previous
estimates of the cluster metallicity \citep[see
  e.g.][]{carretta09,lovisi12}.  In addition, \com\ and MS1 show very
similar values of the Na, Mg and Al abundances.  These elements are
involved in the chemical anomalies usually observed in GCs
\citep{gratton12} and are explained as due to two or more bursts of
star formation in the early phases of the
cluster evolution \citep[see][]{dercole08}. In particular,
given the ranges of values measured in unevolved stars both in NGC
6397 \citep{gratton01,carretta05,pasquini08} and in GCs of similar metallicity
\citep{carretta09}, the [C/Fe], [Mg/Fe] and [Al/Fe] abundance ratios,
as well as the upper limit of [Na/Fe] derived for MS1 suggest with
this object belongs to the first generation of stars
\footnote{\citet{lind09} derive for Turnoff stars in NGC~6397 
temperatures lower than those predicted by the BaSTI isochrone. 
We repeated the analysis of MS1 decreasing $T_{eff}$ by 250 K, in order 
to match the $T_{eff}$ scale by \citet{lind09}. The differences in the 
derived [C/Fe] and [N/Fe] are smaller than 0.2 dex and do not change 
our conclusions about this star.}. In turn, the
observed chemical similarity indicates that this is likely the case
also for \com.

The [N/Fe] upper limit obtained for MS1 is also
consistent with what expected for the first stellar generation
\citep[see][]{carretta05}, while the value measured for
\com\ ([N/Fe]$= +0.53\pm 0.15$ dex) is significantly larger.
Fig. \ref{nh} shows the observed spectrum of \com\ in the region
around the NH band, compared with synthetic spectra calculated with
different values of [N/Fe].  Besides the best-fit synthetic spectrum
(thick solid line), two other spectra are shown: one has been computed
assuming [N/Fe]$=0$ dex, consistent with what expected on the surface
of an unperturbed star and the results obtained for MS1; the other one
has [N/Fe]$=+1.4$ dex, which is the value predicted for the NO-cycle
equilibrium (see Sect. \ref{sec:disc}).  Clearly both these additional
values are incompatible with the measured abundance. As discussed in
Section 5, this provides interesting constraints on the structure and
the nature of this star.

Because of the weakness of the G-band only an upper limit for the C
abundance of \com\ is derived: at the XSHOOTER resolution, we
measure [C/Fe]$<-1.0$ dex. This is compatible with the (more
stringent) limit derived by \citet{sabbi03b} from high-resolution UVES
spectra ([C/Fe]$<-2.0$ dex), which is therefore adopted in the
following discussion. Note that these upper limits are significantly
smaller than any C abundance measured in NGC 6397 stars
\citep[e.g.][]{carretta05}. 

\section{Discussion}
\label{sec:disc}
The abundances of C and N measured for \com\ are incompatible with
the values expected on the surface of an unpeeled MS star.  
Also, its very low [C/Fe] is incompatible with the 
C abundances range observed in the cluster \citep{carretta05}, excluding 
that \com\ is an unpeeled second generation star.
Hence, option $(1)$ discussed in the
Introduction can be ruled out. In order to verify the possibility that
\com\ is, instead, a deeply peeled Turnoff/sub-giant star and to put
new constraints on its nature, we compare the derived C and N
abundances with the chemical gradients predicted in theoretical
stellar models.

We have calculated the evolution of a 0.8 \Msun\ stellar model, from
the pre-MS to the red giant branch phase, using the same code and
input physics of the BaSTI models \citep[see e.g][]{pietr04,pietr06}.
We adopted Z=0.0003, 
Y=0.245, and an $\alpha$-enhanced metal mixture([$\alpha$/Fe]=+0.4), 
which is appropriate for the first generation of
stars in Galactic GCs (in particular, we assume [C/Fe]=0 and [N/Fe]=0,
also consistently with the abundances measured for MS1).  The Turnoff
age of the model is 12 Gyr. This value, however, depends on the
efficiency of atomic diffusion, which can be (partially or totally)
inhibited by additional turbulent mixing (for which an adequate
physical description is still lacking). We therefore calculated models
both with and without atomic diffusion, finding that they are
basically indistinguishable (in the following we therefore presents
only the results obtained from models without atomic diffusion).  Also
the effects of radiative levitation are totally negligible at the
metallicity of NGC 6397, because the radiative acceleration on the C
and N atoms is always smaller than the gravitational acceleration
\citep[see Fig. 3 in][]{richard02} and we therefore do not include
them in our models.

\begin{figure}[h]
\plotone{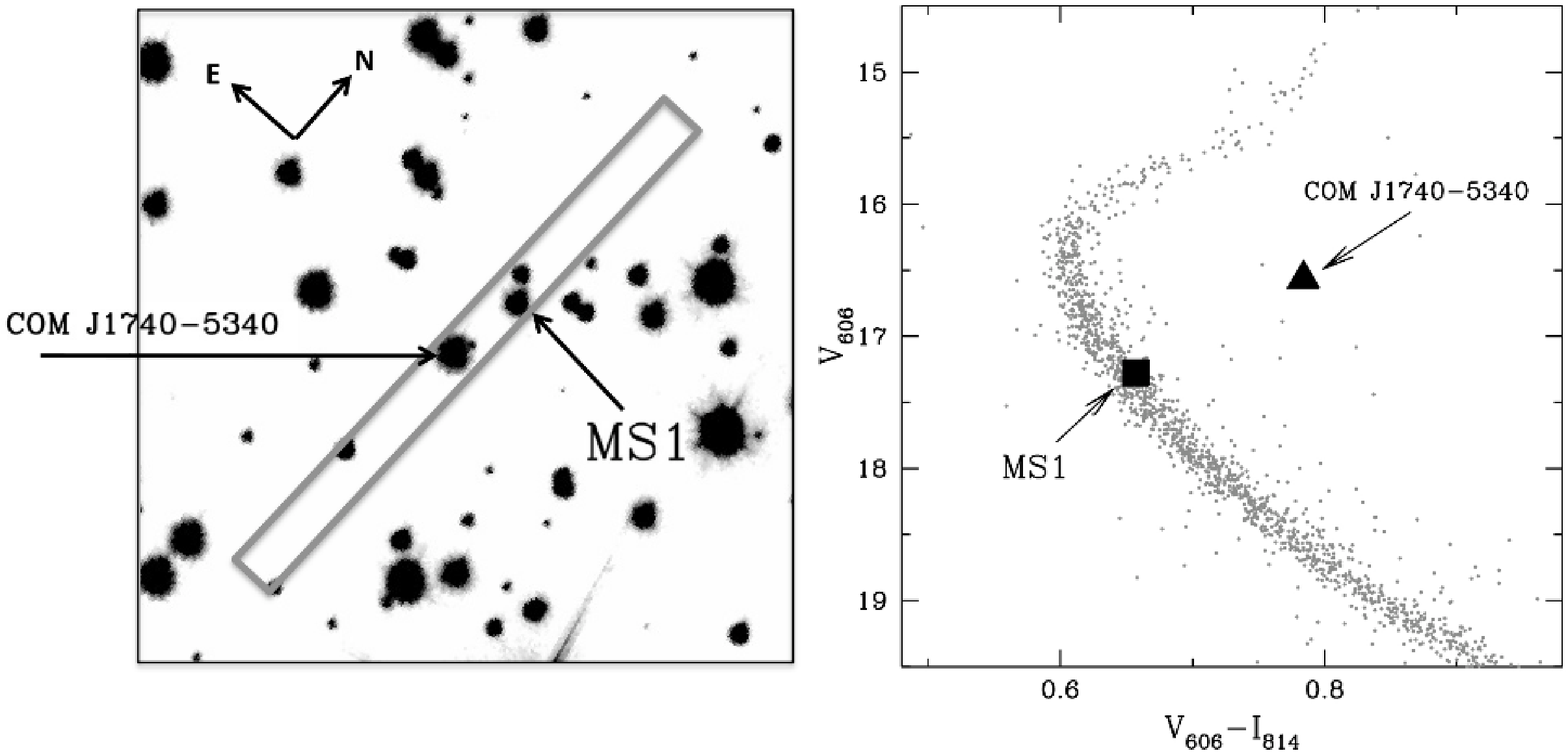}
\caption{ 
{\sl Left panel}: 
HST-ACS archivial $V_{606}$-band image with the position of the adopted XSHOOTER slit 
and the identification of the two targets. 
{\sl Right panel}: 
CMD of the radial (R$<$40 arcsec) region around \com\ 
(Contreras Ramos et al., 2013, in preparation), with marked the position 
of \com\ and MS1.}
\label{loca}
\end{figure}

Fig. \ref{cnm} shows the gradients of the [$^{12}$C/Fe] and
[$^{14}$N/Fe] abundance ratios in the interior of a sub-giant
star\footnote{It is not easy to identify the evolutionary stage of the
  star in the scenario of a peeled star.  Its position in the CMD
  (Fig. \ref{loca}) suggests that the object is a slightly evolved
  star (see also \citealp{burderi02}), but we cannot exclude that it still belongs to
  the MS. In any case, its luminosity seems to exclude that it is a
  giant star.}, as a function of the stellar mass, from the
center to the surface ($M=0.8$ \Msun). 
As apparent, these abundances remain constant along the entire stellar envelope, 
from the surface, down to the radius including half of the total mass. 
The fflat chemical profiles at [N/Fe]$=+1.4$ dex and [C/Fe]$\sim-1$ dex
observed in the very central region ($M\lsim 0.15$ \Msun) are the
consequence of the CNO-burning that occurred in the stellar core
during the MS evolution.  The gradients observed in the intermediate
region (0.15 \Msun$\lsim M\lsim 0.4$ \Msun), instead, are due to the
ongoing hydrogen burning in a thick shell above the inactive core 
(with a mass of $\sim$0.11 \Msun). 
In particular, the most external portion of the shell is mainly
interested by the CN-cycle and therefore shows an increase of $^{14}$N
and a drop of $^{12}$C.  At 0.18 \Msun$\lsim M\lsim 0.38$ \Msun\ the
abundances of $^{12}$C and $^{14}$N reach the equilibrium: C displays
its minimum value ([C/Fe]$=-2.45$ dex) and the N abundance profile
shows a {\sl plateau} at [N/Fe]$=+0.68$ dex. In the innermost portion
of the shell the NO-cycle is active, thus producing a further increase
of both $^{14}$N and $^{12}$C, up to the most central values.

\begin{figure}[h]
\plotone{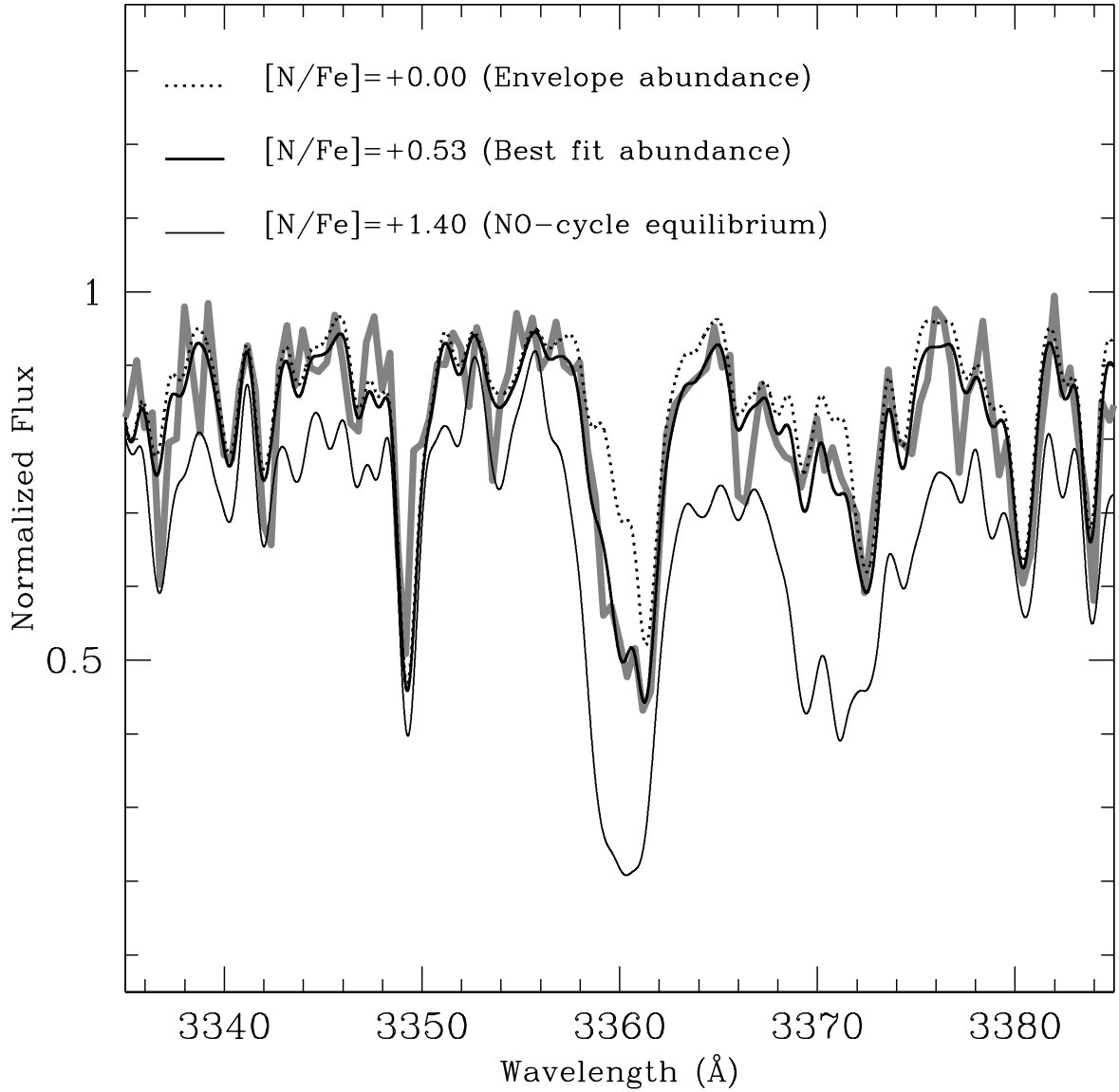}
\caption{Observed spectrum of \com\ (thick grey line) in the
  spectral region around the NH molecular band, 
  with overimposed  synthetic spectra calculated with [N/Fe]=+0.0 
  (corresponding to the
stellar envelope abundance, dotted line), +0.53 (best fit abundance, thick solid line), 
+1.40 (NO-cycle equilibrium abundance, thin solid line).}
\label{nh}
\end{figure}

\begin{figure}[h]
\plotone{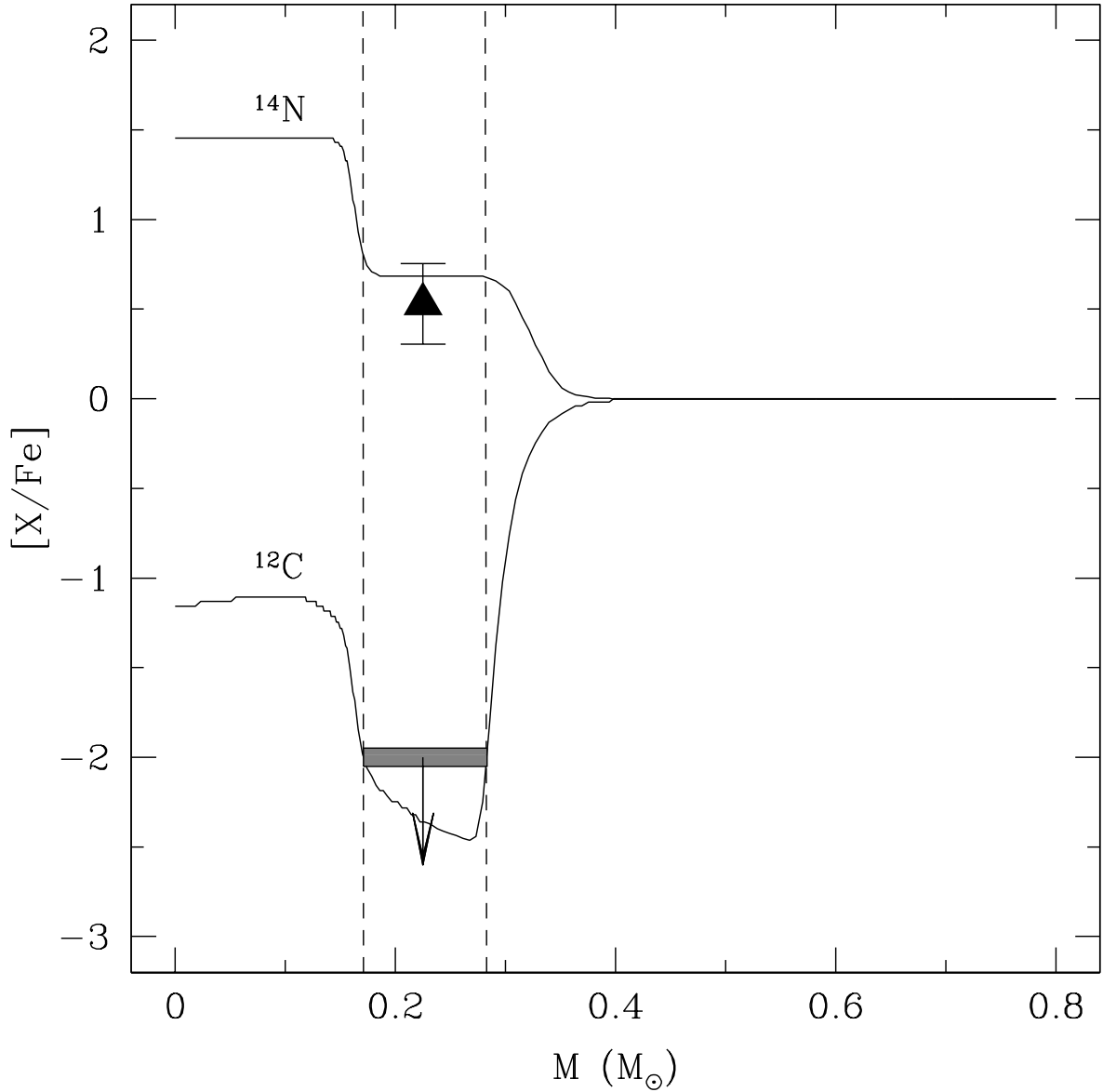}
\caption{
Behavior of [X/Fe] abundance ratio for $^{12}$C and $^{14}$N as a function 
of the mass, for a stellar model of a sub-giant star with M=~0.8 \Msun, 
Z=~0.0003 and no atomic diffusion. 
The black triangle indicates the [N/Fe] measured in \com\, while the grey bar is the upper limit 
for [C/Fe]. The dashed vertical lines mark the mass range defined by the upper limit for [C/Fe].}
\label{cnm}
\end{figure}

The dashed vertical lines in Fig. \ref{cnm} mark the range of mass
profile where the carbon abundance of the stellar model is in
agreement with the upper limit ([C/Fe]$<-2$ dex) derived for
\com\ \citep{sabbi03b}.  In this same mass range
also the N abundance shows a good agreement between the model
prediction and the measured value (black triangle in
Fig. \ref{cnm}). Instead the [N/Fe] ratio observed in \com\ is
incompatible with the abundance ratio predicted in any other region of
the stellar model (consistently with what discussed above; see
Fig. \ref{nh}).  This evidence strongly suggests that \com\ is a star
peeled down to the region where the CN-cycle occurs, as a result of
heavy mass transfer onto the neutron star.

\begin{figure}[h]
\plotone{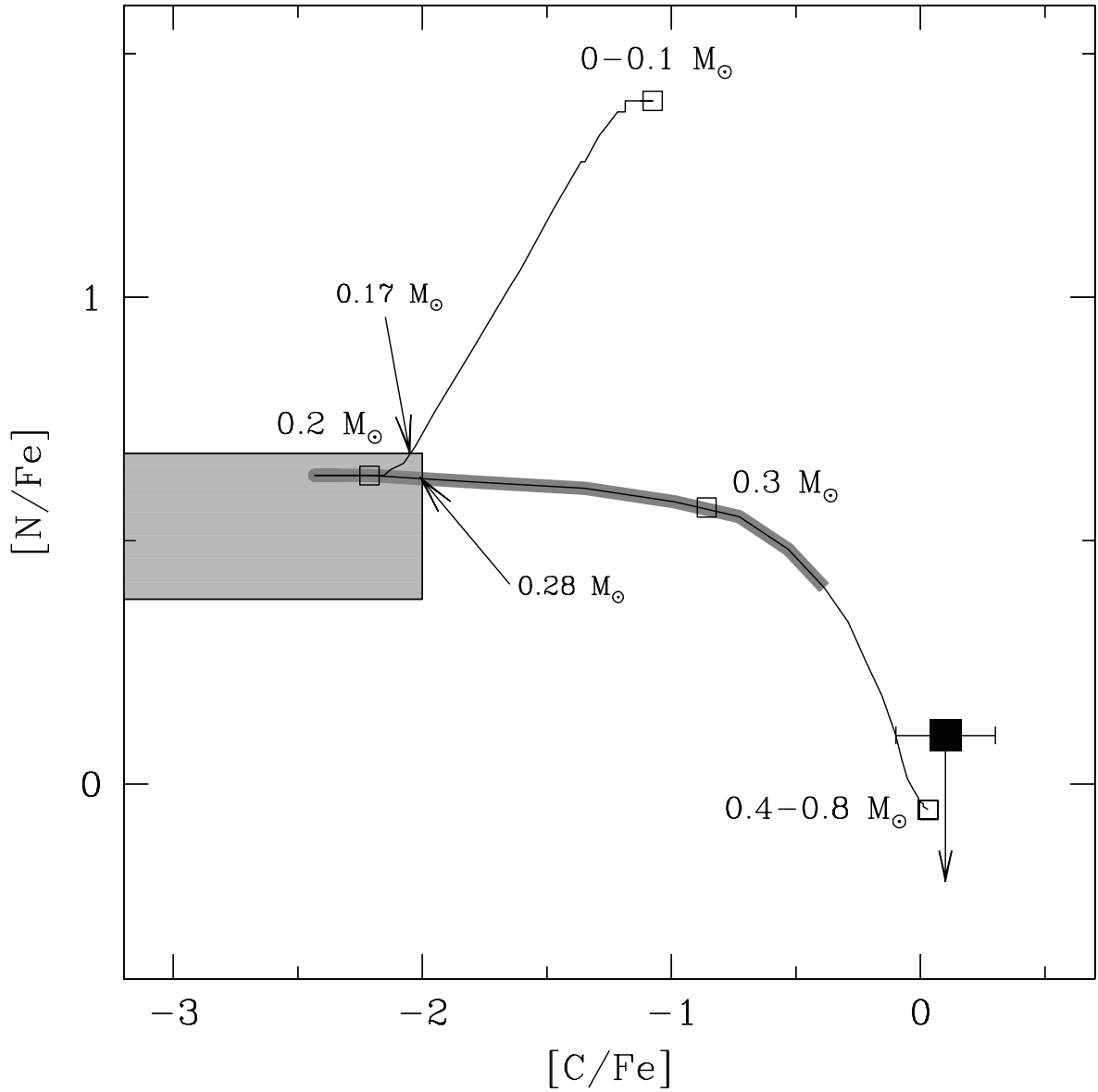}
\caption{
Behavior  of [N/Fe] as a function of [C/Fe] (thin solid lines) for the same stellar model 
shown in Fig.~\ref{cnm}. Empty squares mark the position of the stellar masses 
at step of 0.1 \Msun. The light grey box indicates the mean locus of [C/Fe] and [N/Fe] of 
COM J1740-5340.
The dark grey region indicates the mass range proposed by \citet{ferraro03}. The black square
marks the position of MS1.}
\label{cn}
\end{figure}

The C and N abundances allow us to also identify a reasonable
mass range for the MSP companion. Fig. \ref{cn} shows the behavior of
the discussed stellar model in the [C/Fe]--[N/Fe] plane. The light
grey box indicates the locus corresponding to the abundances
of \com\ (taking into account the quoted uncertainties). Clearly, only
the portion of the stellar model between 0.17 and 0.28
\Msun\ overlaps this region.  This mass range is in very good
agreement with the value (0.22-0.32 \Msun; \citealp{ferraro03})
estimated for \com\ from the binary system mass ratio (inferred from
the radial velocity curve of the companion) and the orbital
inclination angle (inferred from the optical light curve). 
In order to quantify how much the results
depend on the evolutionary phase of the star before the onset of heavy
mass transfer, we repeated the analysis by using models on the MS, at
the base of the Red Giant Branch (RGB), before and after the occurrence of the First
Dredge-Up.  Following the evolution from the Turnoff to the base of
the RGB, the region where the NO-cycle occurs increases in mass, thus
reducing the mass range where the model C and N abundances match the
observed ones. If the MSP companion was a Turnoff star,
\com\ should now have a mass between 0.13 and 0.27 \Msun, while the
range decreases significantly in case of a RGB star: 0.22-0.28
\Msun\ for a star before the First Dredge-Up, and 0.25-0.28 \Msun\ for
a star after the First Dredge-Up.  Note that in all the cases, the
upper mass limit remains basically unchanged, confirming a value
smaller than 0.3 \Msun, even if the upper limit for the C abundance
inferred from the XSHOOTER spectra ([C/Fe]$\sim -1$) is assumed. 
  We finally stress that different assumptions about the initial
  C and N abundances of the star have the effect of rigidly shifting
  the model curves in both the planes shown in Figs. \ref{cnm} and
  \ref{cn}.  
  
The evidence presented here adds an important piece of
information to properly characterize \com.   
The analysis of the C and N surface abundances
provides a diagnostic of the companion mass which is totally
independent of other, commonly used methods, well confirming the
previous estimates. In addition, the chemical patterns observed at the
surface of \com\ solidly confirm that this object is a deeply peeled
star, and can be even used to obtain a quantitave evaluation of 
the amount of mass lost by the star. 
In fact, our analysis indicates that the entire envelope of the
star has been completely removed and the peeling action has
extended down to an interior layer where the CN-cycle approximately
reached the equilibrium
By assuming an initial mass of
$\sim 0.8$ \Msun, we estimate that \com\ has lost $\sim 75\%$ of its
initial mass during the interaction with the pulsar.

\acknowledgements  

We thanks the anonymous referee for his/her suggestions.
This research is part of the project COSMIC-LAB funded by the European Research Council 
(under contract ERC-2010-AdG-267675).

\newpage

\begin{deluxetable}{lll}
\tablecolumns{3} 
\tablewidth{0pc}  
\tablecaption{Chemical abundances measured  for \com\ and MS1.}
\tablehead{\colhead{Ratio} & \colhead{COM J1740-5340}& \colhead{MS1}}
\startdata 
\hline	   
${\rm [Fe/H]}$     & $-2.00 \pm 0.12$ & $-1.93 \pm 0.18$ \\
${\rm [Mg/Fe]}$    &  $+0.38 \pm 0.13$ &  $+0.30\pm0.15$ \\
${\rm [Al/Fe]}$    &  $+0.31 \pm 0.14$ &  $+0.35\pm0.20$ \\
${\rm [Na/Fe]}$    &  $<0.00$          &  $< 0.15$       \\
\hline
${\rm [N/Fe]}$     &  $+0.53 \pm 0.15$  &  $<+0.10$ \\
${\rm [C/Fe]}$     &  $<-1.0$          &  $+0.10\pm 0.20$   \\
\hline
\enddata 
\tablecomments{Reference solar abundances are by \citet{gs98}.}
\end{deluxetable}


\begin{thebibliography}{}
\bibitem[Andrievsky et al.(2008)]{andr08}
Andrievsky, S. M., Spite, M., Korotin, S. A., Spite, F., Bonifacio, P., 
Cayrel, R., Hill, V., \& Francois, P., 2008, A\&A, 481, 481
\bibitem[Bailyn(1995)]{bai95} Bailyn, C.~D.\ 1995, \araa, 33, 133 
\bibitem[Bellazzini et al.(1995)]{bella95} Bellazzini, M., 
Pasquali, A., Federici, L., Ferraro, F.~R., 
\& Pecci, F.~F.\ 1995, \apj, 439, 687
\bibitem[Burderi et al.(2002)]{burderi02}
Burderi, L., D'Antona, F., \& Burgay, M., 2002, ApJ, 574, 325
\bibitem[Carretta et al.(2005)]{carretta05}
Carretta, E., Gratton, R. G., Lucatello, S., Bragaglia, A., \& Bonifacio, P., 
2005, A\&A, 433, 597
\bibitem[Carretta et al.(2009)]{carretta09}
Carretta, E., Bragaglia, A., Gratton, R. G., D'Orazi, V., \& Lucatello, S., 2009, 
A\&A, 508, 695
\bibitem[Castelli \& Kurucz(2004)]{castelli04}
Castelli, F. \& Kurucz, R. L., 2004, arXiv:astro-ph/0405087v1
\bibitem[D'Amico et al.(2001a)]{damico01a}
D'Amico, N., Lyne, A. G., Manchester, R. N., Possenti, A., \& Camilo, F., 2001, ApJ, 548L, 171
\bibitem[D'Amico et al.(2001b)]{damico01b}
D'Amico, N., Possenti, A., Manchester, R. N., Sarkissian, J. Lyne, A. G., \& Camilo, F., 2001, 561L, 89
\bibitem[D'Ercole et al.(2008)]{dercole08} D'Ercole, A., Vesperini,
  E., D'Antona, F., McMillan, S.~L.~W., \& Recchi, S.\ 2008, \mnras,
  391, 825
\bibitem[Ergma \& Sarna(2003)]{ergma03}
Ergma, E., \& Sarna, M. J., 2003, A\&A, 399, 237
\bibitem[Ferraro et  al.(1995)]{ferraro95} Ferraro, F.~R., Fusi Pecci, F., \& Bellazzini, M.\ 1995, \aap, 294, 80 
\bibitem[Ferraro et al.(1999)]{f99}
Ferraro, F. R., et al., 1999, AJ, 118, 1738
\bibitem[Ferraro et al.(2001)]{ferraro01}
Ferraro, F. R., Possenti, A., D'Amico, N. \& Sabbi, E., 2001, ApJ, 561L, 93
\bibitem[Ferraro et al.(2003)]{ferraro03}
Ferraro, F. R., Sabbi, E., Gratton, R., Possenti, A., D'Amico, N., 
\& Camilo, F., 2003, ApJ, 589L, 41
\bibitem[Ferraro et al.(2006)]{ferraro06} Ferraro, F.~R., Sabbi, 
E., Gratton, R., et al.\ 2006, \apjl, 647, L53 
\bibitem[Ferraro et al.(2009)]{ferraro09} Ferraro, F.~R., 
Beccari, G., Dalessandro, E., et al.\ 2009, \nat, 462, 1028 
\bibitem[Ferraro et al.(2012)]{ferraro12} Ferraro, F.~R., 
Lanzoni, B., Dalessandro, E., et al.\ 2012, \nat, 492, 393 
\bibitem[Gratton et al.(2001)]{gratton01}
Gratton, R. G., et al., 2001, A\&A, 369, 87
\bibitem[Gratton et al.(2012)]{gratton12}
Gratton, R. G., Carretta, E., \& Bragalia, A., 2012, A\&AR, 20, 50
\bibitem[Grevesse \& Sauval(1998)]{gs98}
Grevesse, N., \& Sauval, A. J., 1998, Space Science Reviews, 85, 161
\bibitem[Grindlay et al.(2001)]{grindlay01} Grindlay, J.~E., Heinke,
  C.~O., Edmonds, P.~D., Murray, S.~S., \& Cool, A.~M.\ 2001, \apjl,
  563, L53
\bibitem[Grindlay et al.(2002)]{grindlay02} Grindlay, J.~E., Camilo,
  F., Heinke, C.~O., et al.\ 2002, \apj, 581, 470
\bibitem[Kaluzny et al.(2003)]{kaluzny03}
Kaluzny, J., Rucinski, S. M., \& Thompson, I. B., 2003, AJ, 125, 1546
\bibitem[Lind et al.(2009)]{lind09}
Lind, K., Primas, F., Charbonnel, C., Grundhal, F., \& Asplund, M., 
2009, A\&A, 503, 545
\bibitem[Lind et al.(2010)]{lind}
Lind, K., Asplund, M., Barklem, P. S., \& Belyaev, A. K., 2011, A\&A, 528, 103
\bibitem[Lovisi et al.(2012)]{lovisi12}
Lovisi, L., Mucciarelli, A., Lanzoni, B., Ferraro, F. R., Gratton, R. G., Dalessandro, E., 
\& Contreras Ramos, R., 2012, ApJ, 754, 91
\bibitem[Lucatello \& Gratton(2003)]{lucatello03}
Lucatello, S., \& Gratton, R. G., 2003, A\&A, 406, 691
\bibitem[Mucciarelli, Salaris \& Bonifacio(2012)]{m12}
Mucciarelli, A., Salaris, M., \& Bonifacio, P., 2012, MNRAS, 419, 2195
\bibitem[Mucciarelli et al.(2012)]{m12b}
Mucciarelli, A., Bellazzini, M., Ibata, R., Merle, T., Chapman, S. C. Dalessandro, E., 
\& Sollima, A., 2012, MNRAS, 426, 2889
\bibitem[Neckel \& Labs(1984)]{neckel}
Neckel, H., \& Labs, D., 1984, SoPh, 90, 205
\bibitem[Orosz \& van Kerkwijk(2003)]{orosz}
Orosz, J. A. \& van Kerkwijk, M. H., 2003, A\&A, 397, 237
\bibitem[Pasquini et al.(2008)]{pasquini08}
Pasquini, L., Ecuvillon, A., Bonifacio, P., \& Wolff, B., 2008, A\&A, 489, 315
\bibitem[Pietrinferni et al.(2004)]{pietr04}
Pietrinferni, A., Cassisi, S., Salaris, M., \& Castelli, F. 2004, ApJ, 612, 168
\bibitem[Pietrinferni et al.(2006)]{pietr06}
Pietrinferni, A., Cassisi, S., Salaris, M., \& Castelli, F. 2006, ApJ, 642, 797
\bibitem[Possenti(2002)]{possenti}
Possenti, A., Proceedings of the 270. WE-Heraeus Seminar on Neutron Stars, Pulsars, and Supernova Remnants. 
MPE Report 278. Edited by W. Becker, H. Lesch, and J. Trümper. Garching bei München: Max-Plank-Institut für 
extraterrestrische Physik, 2002., p.183
\bibitem[Richard, Michaud \& Richer(2002)]{richard02}
Richard, O., Michaud, G., \& Richer, J., 2002, ApJ, 580, 1100
\bibitem[Sabbi et al.(2003a)]{sabbi03a}
Sabbi, E, Gratton, R. G.,  Ferraro, F. R., Bragaglia, A., Possenti, A., 
D'Amico, N., \& Camilo, F., 2003a, ApJ, 589, L41
\bibitem[Sabbi et al.(2003b)]{sabbi03b}
Sabbi, E, Gratton, R. G., Bragaglia, A., Ferraro, F. R., Possenti, A., 
Camilo, F., \& D'Amico, N., 2003b, A\&A, 412, 829
\bibitem[Sbordone et al.(2004)]{sbordone}
Sbordone, L., Bonifacio, P., Castelli, F., \& Kurucz, R. L., 2004, MSAIS, 5, 93
\end{thebibliography}
\end{document}